\documentclass[a4paper,11pt]{article}
\usepackage{pos}

\usepackage{siunitx}
\usepackage{lineno}

\usepackage{xcolor}

\newcommand{\refeq}[1]{Eq.~(\ref{#1})}
\newcommand{\refeqs}[2]{Eqs.~(\ref{#1})~and~(\ref{#2})}

\newcommand{\reffig}[1]{Fig.~\ref{#1}}

\newcommand{\refsec}[1]{Section~\ref{#1}}

\title{Radio measurements of air showers with the IceCube-Gen2 surface prototype station at the Pierre Auger Observatory}
\ShortTitle{Radio measurements with IceCube station at Auger}

\manuallySeparateAuthors
\author*[a]{Stef Verpoest}
\author[1]{for the IceCube-Gen2}
\author{ and }
\author[b,2]{Pierre Auger}
\author{ Collaborations}

\affiliation[a]{Bartol Research Institute, Department of Physics and Astronomy, University of Delaware,\\
  Sharp Lab, 104 The Green, Newark DE, 19716, United States of America}
\affiliation[b]{Observatorio Pierre Auger, Av. San Martín Norte 304, 5613 Malargüe, Argentina}

\emailAdd{stef.verpoest@icecube.wisc.edu}
\emailAdd{analysis@icecube.wisc.edu}
\emailAdd{spokerspersons@auger.org}

\note{Full author list at \url{https://icecube.wisc.edu/collaboration/authors}.}
\note{Full author list at \url{https://www.auger.org/archive/authors_2026_06.html}.}

\abstract{The design of the IceCube-Gen2 observatory, proposed as a next-generation extension of IceCube, includes a surface array consisting of scintillators and radio antennas. In addition to several such detectors already deployed at IceCube’s surface array at the South Pole, a complete prototype station including three SKALA antennas has been operating at the Pierre Auger Observatory for several years. This setup has been used to successfully observe radio signals from air showers, demonstrated through coincident observations with the Auger Surface Detector. In this contribution, we present an updated analysis of these radio signals, and compare them to CoREAS simulations using the reconstruction from the Auger Surface Detector as input. }

\FullConference{11th International Workshop on Acoustic and Radio EeV Neutrino Detection Activities (ARENA2026)\\
8-11 June 2026\\
Karlsruhe, Germany\\}

\begin{document}
\maketitle

\section{Introduction}

The IceCube Neutrino Observatory~\cite{IceCube:2016zyt} at the South Pole includes a surface
air-shower detector, IceTop~\cite{IceCube:2012nn}, which is being enhanced with scintillation detectors and radio
antennas. The planned next-generation extension, IceCube-Gen2~\cite{IceCube-Gen2:2020qha}, also
foresees a surface array based on these detector types~\cite{IceCube-Gen2:2021aek}. Prototype
stations have been deployed at the South Pole~\cite{IceCube:2021epf, IceCube:2023rrl} and at
the Pierre Auger Observatory~\cite{PierreAuger:2015eyc} in Argentina, each featuring eight scintillator panels and three radio
antennas of the SKALA~v2 type~\cite{7297231}. The readout of the radio signals is triggered when six scintillator record a coincident signal. The station at Auger is located within the SD-433
area~\cite{PierreAuger:2021tmd}, the most densely instrumented part of the surface detector array.
The successful detection of air-shower radio signals with this station, identified through
coincident observations with the Auger SD-433 array, has been previously
reported~\cite{Verpoest:2024fi, IceCube-Gen2:2025aym}. In this contribution, we present an updated
analysis including new data, and include a systematic comparison of the measured radio
waveforms against CoREAS~\cite{Huege:2013vt} simulations using the SD-433 shower reconstruction
as input.

\section{Identified Air Shower Events}\label{sec:events}

The identification of air showers observed with the radio antennas of the prototype
station, in coincidence with the Auger SD-433 array, was presented in earlier
work~\cite{Verpoest:2024fi, IceCube-Gen2:2025aym}. Here, we briefly summarize the
selection procedure and present an updated sample. Starting from the
scintillator-triggered events, the radio waveforms are processed, including corrections
for DAQ artifacts, bandpass filtering to the \SI{110}{\mega\Hz} to \SI{185}{\mega\Hz}
band, and the application of a frequency-weighting scheme to further suppress
narrow-band background~\cite{IceCube:2021qnf}. Using the signal-to-noise ratio (SNR)
distributions obtained from background (fixed-rate trigger) waveforms, we select events
with at least three high-SNR waveforms. For each selected event, a directional
reconstruction is performed by fitting a planar shower front to the signal peak times in
the antennas. The candidates are then matched to events reconstructed by the Auger
SD-433 array~\cite{PierreAuger:2023dju} based on their trigger times and reconstructed
arrival directions.

The present sample uses data recorded at a sampling rate of \SI{800}{Msps}, collected
between October 2023 and March 2024, and between April 2025 and October 2025.
After removing periods of downtime due to hardware and software issues, this corresponds
to a total runtime of 293.6 days. In total, 215 events with radio emission matching an
event reconstructed by SD-433 were identified. Compared to the earlier analysis, an
improved treatment of DAQ artifacts in the radio waveforms increased the rate of
identified events by about 20\%.

\begin{figure}
  \centering
  \includegraphics[width=0.5\linewidth]{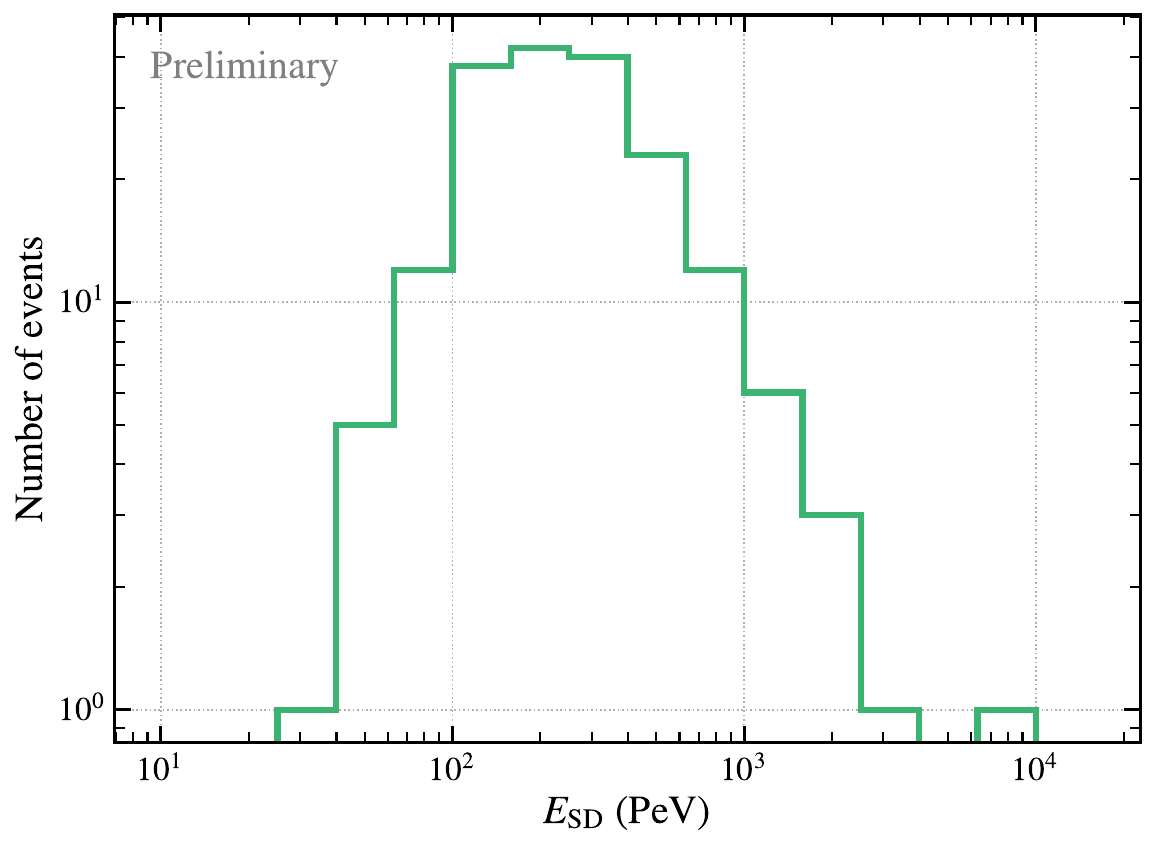}\hfill\includegraphics[width=0.45\linewidth]{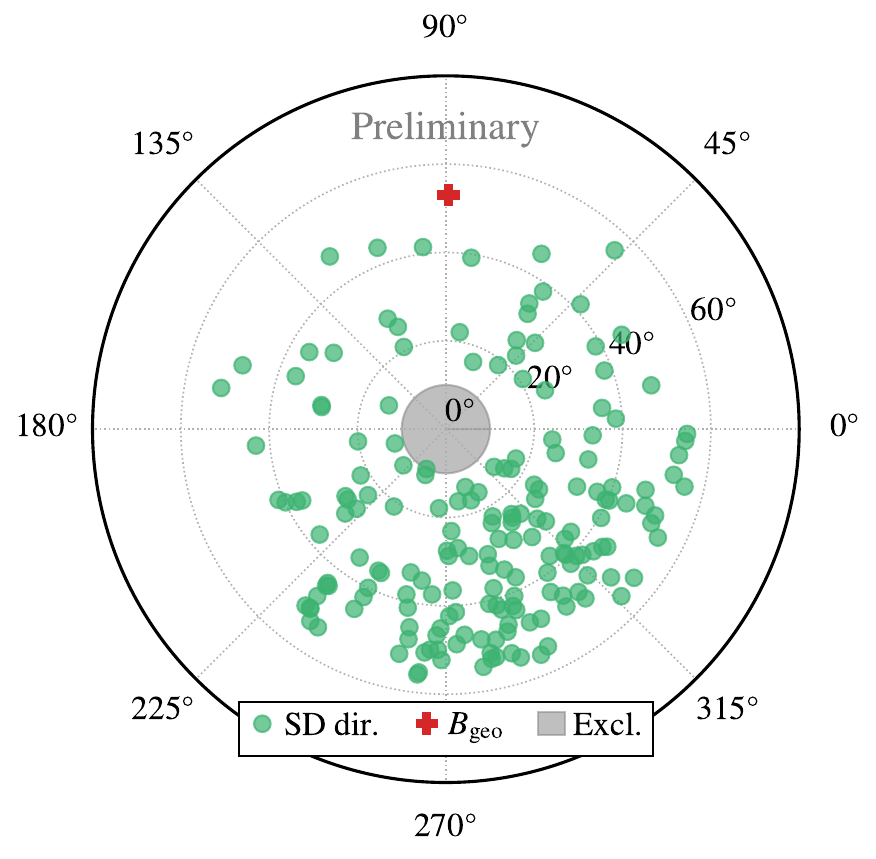}
  \caption{Properties of the identified air-shower events, as reconstructed by the
  Auger SD-433 array. Left: Distribution of the reconstructed shower energies (excluding
  31 events without a successful SD-433 energy reconstruction). Right: Polar plot of the
  reconstructed arrival directions; the direction of the geomagnetic field is indicated,
  and the region around the zenith is excluded from the analysis.}
  \label{fig:sample}
\end{figure}

The reconstructed energies and arrival directions of the identified events are shown in
\reffig{fig:sample}, using values from the SD-433 reconstruction. Of the 215 matched
events, 31 lack a successful SD-433 energy reconstruction and are excluded from the figures and further analysis; the remaining events span reconstructed energies from \SI{33}{\peta\eV} to \SI{7}{\exa\eV}.
The arrival directions show the expected tendency towards larger angles to the geomagnetic
field. The region around the zenith is excluded from the selection, as some residual DAQ
artifacts can mimic vertical air showers.

\section{Data--Monte Carlo Comparison}

The sample of identified air showers is used to study the agreement between the observed
radio signals and the expectations from simulations. For each event, dedicated simulations
are produced, and observables are calculated that serve as proxies for the polarization of
the radio signal. The method and results are detailed in the following. Before calculating any observables, all waveforms are upsampled by a factor 4 to \SI{3.2}{GSps} by zero-padding in the frequency domain.

\subsection{Goal and Method}

The goal is to compare the properties of the simulated and observed radio signals. To this
end, we compare the relative strength of the signals in the two polarization channels of
each antenna. In this way, the comparison is largely insensitive to the missing absolute
signal-chain calibration and to possible biases in the SD-433 energy estimate
$E_\mathrm{SD}$, while still providing a measure of the polarization of the electric field
at each antenna.

We introduce the observable $Q$, comparing the signal power in the two channels of an
antenna,
\begin{equation}\label{eq:Q}
    Q = \frac{S_0 - S_1}{S_0 + S_1},
\end{equation}
where $S_i = P_i^{\rm signal} - P_i^{\rm noise}$ is the noise-subtracted signal power in
channel $i$, and the power is computed as
\begin{equation}
    P_i = \sum_{t = t_{\rm peak} - w}^{t_{\rm peak} + w} V_i(t)^2
\end{equation}
in a window of half-width $w$ around the signal peak. The value for $w$ used here is 60 bins for the upsampled waveforms or \SI{18.75}{\nano\s}.
The peak time is determined from the Hilbert envelopes $H_i(t)$ of the two channels as
$t_{\rm peak} = \underset{t}{\mathrm{argmax}} \left[ H_0(t)^2 + H_1(t)^2 \right]$.
$P_i^{\rm noise}$ is evaluated by splitting the waveform into chunks of the same length as the signal window, excluding the identified signal window, and taking the median. By
construction, $Q = +1$ ($-1$) if all power is contained in channel~0 (channel~1). The
observable is similar to the normalized Stokes parameter $Q/I$, but defined in the antenna frame directly from the measured voltages rather than from the unfolded electric-field components.

\begin{figure}
    \centering
    \includegraphics[width=0.46\linewidth]{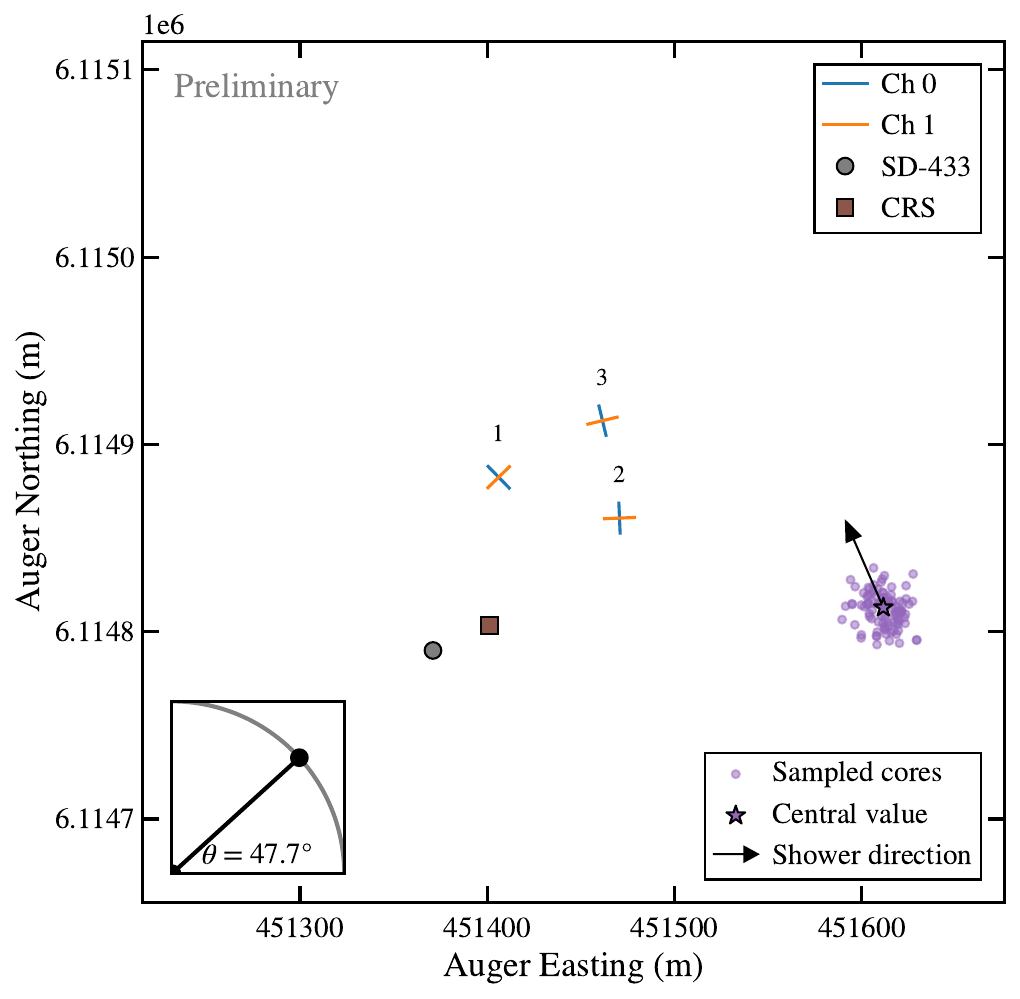}\hfill\includegraphics[width=0.49\linewidth]{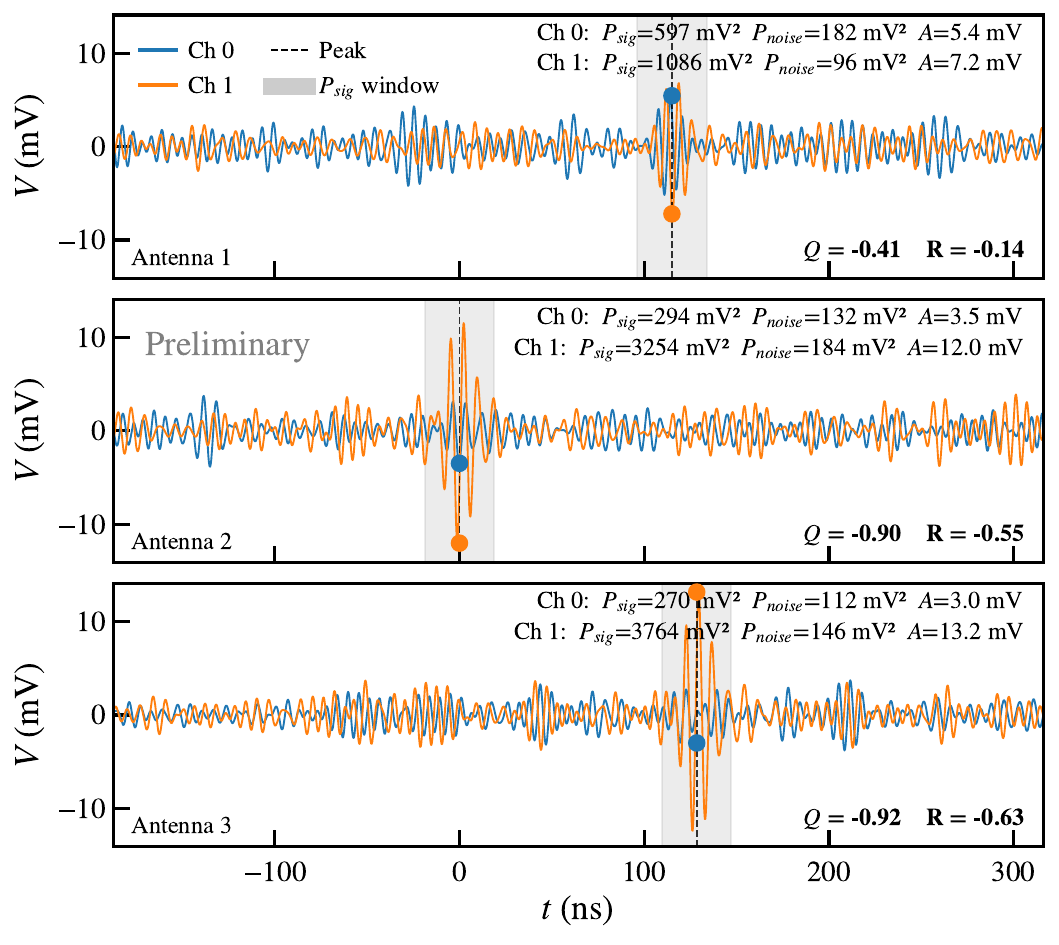}
    \caption{Illustration of the analysis for an example event. Left: The shower core
    reconstructed by SD-433, together with 100 core positions sampled according to the
    SD-433 reconstruction uncertainty (see \refsec{sec:sims}), relative to the antenna
    positions. Right: Measured waveforms of one antenna used to compute the observables
    $Q$ (\refeq{eq:Q}) and $R$ (\refeq{eq:R}), with the signal window indicated.}
    \label{fig:example}
\end{figure}

We define an analogous observable $R$ based on the amplitudes of the Hilbert envelopes,
\begin{equation}\label{eq:R}
    R = \frac{A_0 - A_1}{A_0 + A_1}, \qquad A_i = \underset{t}{\mathrm{max}}\left(H_i(t)\right).
\end{equation}
An example of the calculation is shown in \reffig{fig:example} for a measured event.

\begin{figure}
    \centering
    \includegraphics[width=0.49\linewidth]{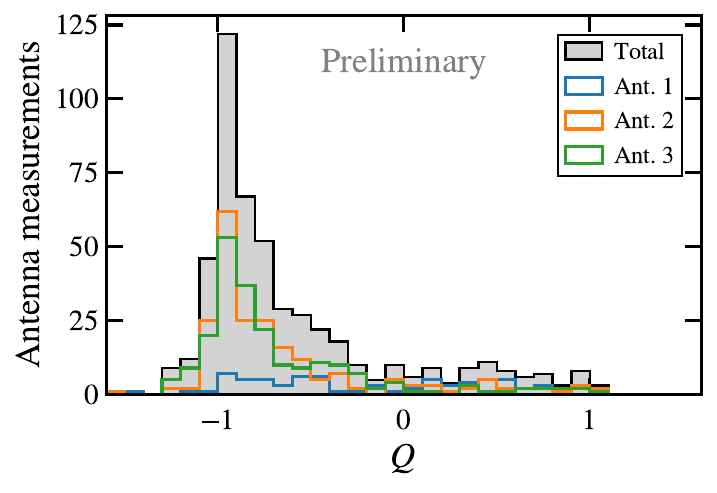}\hfill\includegraphics[width=0.494\linewidth]{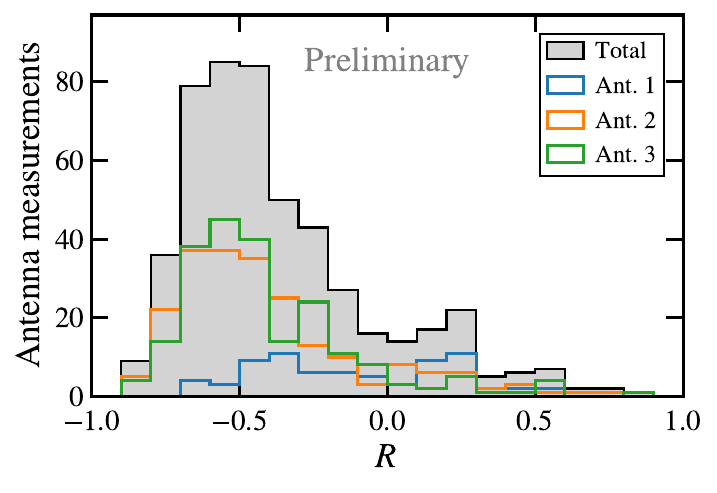}
    \caption{Distributions of the observables $Q$ (left) and $R$ (right) in data, calculated per antenna for the identified air-shower events.}
    \label{fig:data_dists}
\end{figure}

For each identified air-shower event, the signal peaks are identified and $Q$ and $R$ are
computed per antenna. The resulting distributions are shown in
\reffig{fig:data_dists}. The tendency for antennas~2 and~3 towards negative values is a result from their orientation
which points roughly north, placing most of the signal in channel~1 for the majority of events.
Antenna~1 is rotated by about $45^\circ$ from this direction, resulting in more comparable
signals in the two channels. Antenna~1 also contributes fewer events to the distributions, as it was
non-operational during part of the data-taking period.

\subsection{Simulations}\label{sec:sims}

To compare the measurements with expectations, a library of air-shower simulations is
produced for each identified event. The showers are simulated with CORSIKA~\cite{Heck:1998vt}
and the radio emission with CoREAS~\cite{Huege:2013vt}, using the energy and direction from
the SD-433 reconstruction as input, together with the geomagnetic field, observation level,
and measured antenna positions of the Auger site. The Malargüe October atmosphere is used
throughout; the use of a fixed atmosphere is a known limitation, and simulating the
atmosphere specific to the time of each event is left for future work. The simulated radio
signals are propagated through a simulation of the antenna and electronics response as
implemented in the IceTray software framework~\cite{IceCube:2022dcd}.

The impact of the main sources of uncertainty are estimated based on dedicated simulation sets:
\begin{itemize}
    \item \textbf{Longitudinal development:} for each event, 25 showers with proton
    primaries and 15 with iron primaries are simulated, covering the phase space of the
    longitudinal shower development, most importantly the position of the shower maximum
    $X_{\rm max}$.
    \item \textbf{Core position:} in addition to the central SD-433 value, 100 core
    positions are sampled following the SD-433 reconstruction uncertainty of each event
    (an example of the sampled cores is shown in the left panel of \reffig{fig:example}).
    \item \textbf{Background noise:} the simulation is repeated while injecting 100
    different measured background waveforms, randomly selected within the same day as the actual event.
    \item \textbf{Antenna orientation:} the antenna orientations are perturbed 100 times
    following a Gaussian of $1^\circ$ width. While differential-GPS measurements indicate a
    precision of $0.5^\circ$ for the antenna alignment, we use a slightly more conservative value to guard against possible unknown biases or systematic uncertainties.
\end{itemize}
Uncertainties related to the calibration of the signal chain are not yet included.

The central value of each observable is taken from the noise-free simulations, as the
median over the $X_{\rm max}$ distribution using the central core position and antenna
orientation. The associated uncertainties are derived from the variation sets: a min--max
range for $X_{\rm max}$, and 68\% intervals for the core position, antenna orientation, and
background noise. The median shift of the observables induced by noise is also taken into
account. For most events, the noise variability is by far the dominant uncertainty.

\subsection{Results}

\begin{figure}
    \centering
    \includegraphics[width=0.5\linewidth]{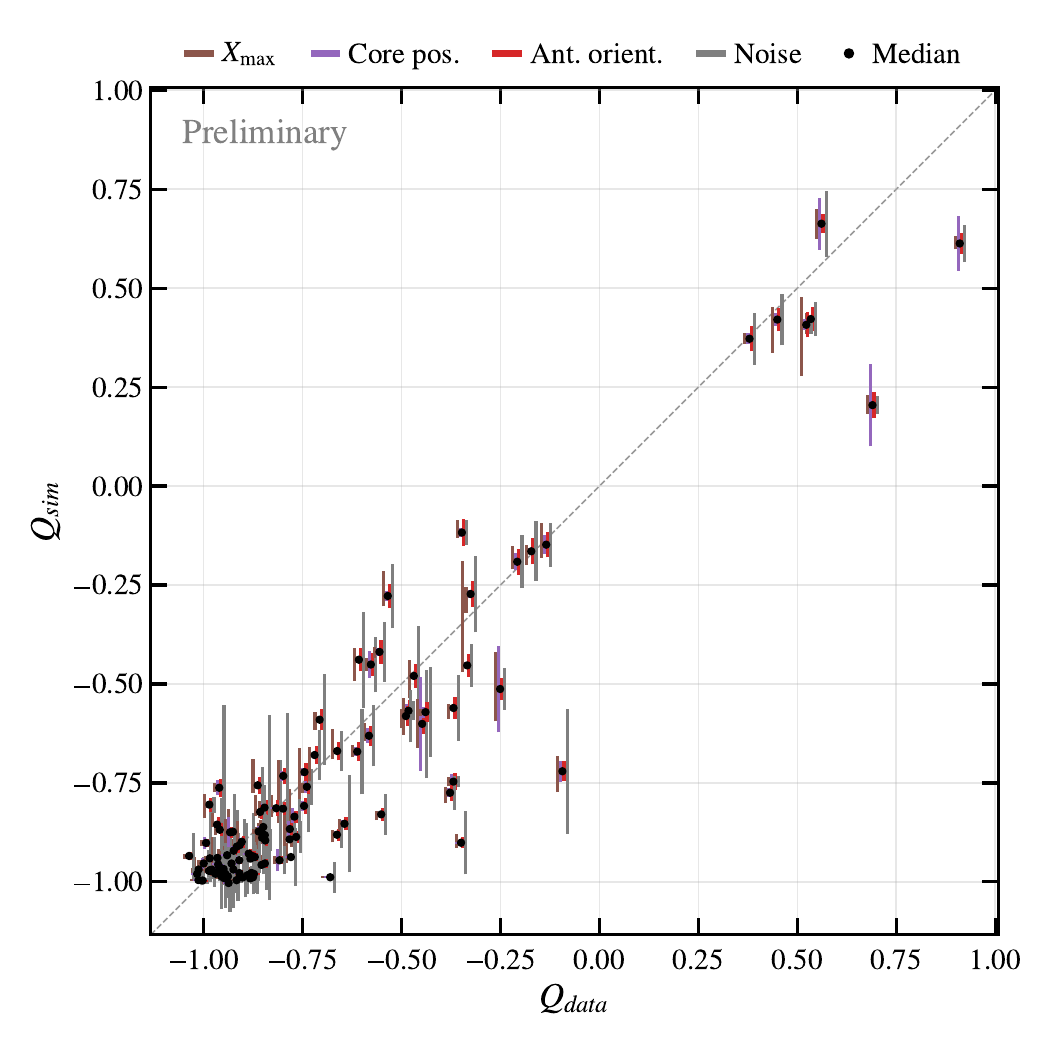}\hfill\includegraphics[width=0.5\linewidth]{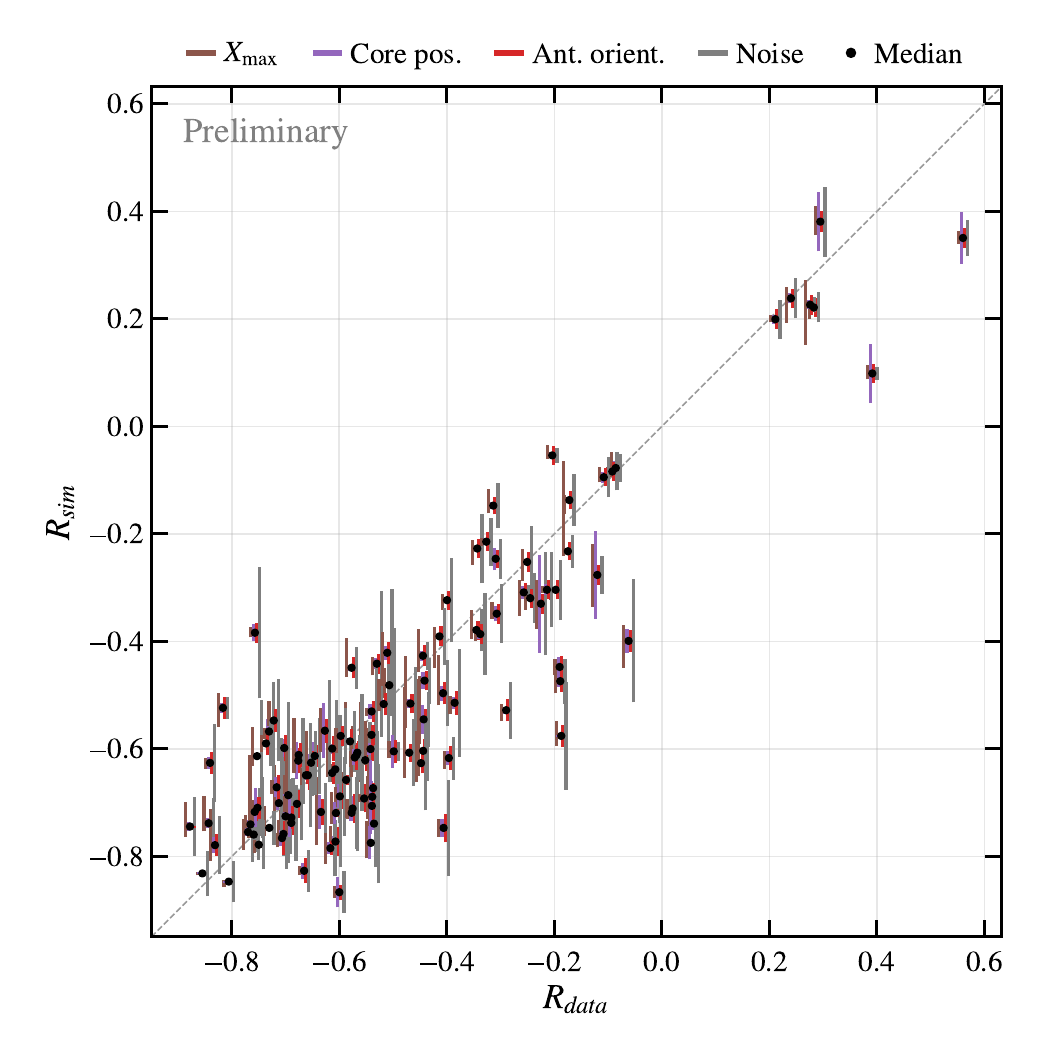}
    \caption{Comparison of the observables $Q$ (left) and $R$ (right) measured in data and
    predicted by simulations. Error bars are shown separately for the different sources of uncertainty considered (see \refsec{sec:sims} for details). Quality cuts were applied to select observations not dominated by background noise.}
    \label{fig:scatter}
\end{figure}

Simulations were produced for all identified air showers. The observables computed in data
and simulation are compared in \reffig{fig:scatter}. Quality cuts are applied to the
simulations, matching those used for the data (per-channel SNR and directional-reconstruction
requirements), together with an additional antenna-level cut on an SNR observable defined based on the total signal and noise power summed over both channels,
\begin{equation}
    \mathrm{SNR}_P = \frac{P_0^{\rm signal} + P_1^{\rm signal}}{P_0^{\rm noise} + P_1^{\rm noise}},
\end{equation}
ensuring that the signal is not dominated by noise and that the peak is correctly
identified (\mbox{$\mathrm{SNR}_P > 4$}). After these cuts, 115 antenna measurements from 70 unique events remain. The
points generally lie along the diagonal, indicating agreement between data and simulation.

\begin{figure}
    \centering
    \includegraphics[width=\linewidth]{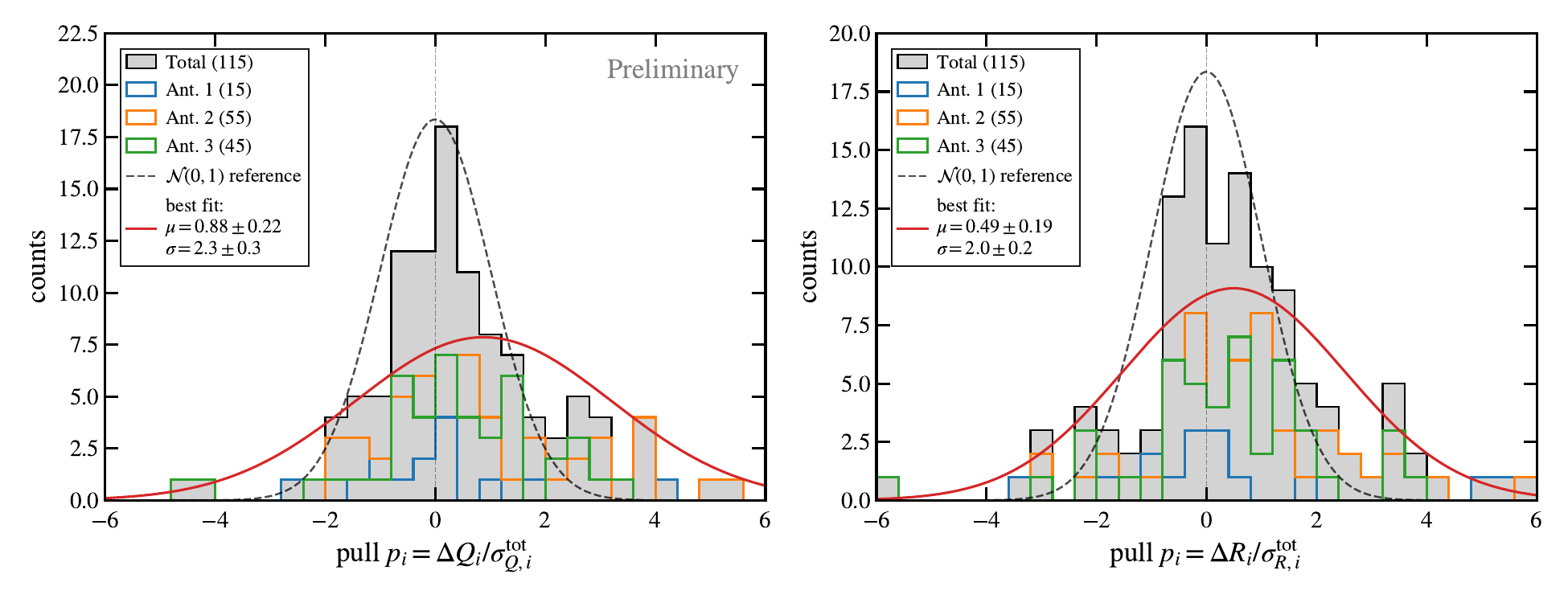}
    \caption{Pull distributions of the observables $Q$ (left) and $R$ (right), shown for the
    three antennas. The pull is the difference between data and simulation divided by the
    total uncertainty. A consistent description within the uncertainties would yield a
    standard normal distribution (indicated).}
    \label{fig:pulls}
\end{figure}

A more condensed view is provided by the pull distributions in \reffig{fig:pulls}, defined
as the difference between the observable in data and simulation divided by the total
uncertainty, with the individual contributions added in quadrature,
$\sigma_{\rm tot}^2 = \sigma_{X_{\rm max}}^2 + \sigma_{\rm core}^2 + \sigma_{\rm orient}^2 + \sigma_{\rm noise}^2$.
For a consistent description of the data within the estimated uncertainties, the
distribution is expected to be normal, which holds for the bulk of the antenna
measurements. However, outliers at large values are present, and an unbinned
maximum-likelihood fit indicates a possible bias ($\mu_Q = 0.88 \pm 0.22$,
$\mu_R = 0.49 \pm 0.19$) and an underestimation of the uncertainties
($\sigma_Q = 2.3 \pm 0.3$, $\sigma_R = 2.0 \pm 0.2$). We verified that these
outliers are not removed by applying stricter SD-433 quality cuts, which additionally
restrict the sample to zenith angles below $45^\circ$. With these cuts, 62 data points
remain, yielding $\mu_Q = 0.59 \pm 0.23$, $\sigma_Q = 1.8 \pm 0.2$ and
$\mu_R = 0.29 \pm 0.22$, $\sigma_R = 1.8 \pm 0.3$.

We conclude that for the majority of the events, the polarization of the air-shower signals is
well described by the simulations, and that our simulated signal chain provides a reasonable
description of reality. Outlier events and the best-fit width of the pull distribution being larger than unity may result from various effects not yet included in our study, such as atmospheric effects and signal-chain calibration
uncertainties. These aspects will be studied in more detail in future work.

\section{Conclusion}

A prototype station of the IceCube-Gen2 surface array has been operating at the Pierre Auger
Observatory for several years, providing a growing sample of air showers observed with its radio antennas, and recent improvements in the low-level data processing have increased the
efficiency of the event identification. For the updated sample of identified air showers presented in this work, a dedicated simulation library was produced, including variations describing several sources of possible disagreement between data and simulations: the
longitudinal shower development, the variability of the background noise, the shower core
position, and the antenna orientation.

Using two observables that serve as proxies for the polarization of the electric field at the
antennas (\refeqs{eq:Q}{eq:R}), we performed a systematic comparison between data and simulation. We find reasonable
overall agreement, albeit with possible indications of an incomplete uncertainty budget. For a
more complete understanding of these results, follow-up studies of the impact of atmospheric
conditions and of the uncertainties in the antenna model and signal-chain calibration will be
performed.

\bibliographystyle{ICRC}
\bibliography{bibliography}

\section*{Acknowledgments}
We thank Eloy de Lera Acedo and Quentin Gueuning for their support regarding the SKALA~v2 antennas.
This project benefited from funding provided by the European Research Council (ERC) and the U.S.~National Science Foundation (NSF).
Further acknowledgments are included with the full author lists at: \url{https://www.auger.org/archive/authors_2026_06.html} and \url{https://icecube.wisc.edu/collaboration/authors/}.

\end{document}